\documentclass{iopart}
\usepackage{iopams}
\usepackage[english]{babel}
\usepackage{cite} %%%%%%
\usepackage{textcomp}
\usepackage{graphics}
\usepackage{graphicx}
\usepackage{anysize}
\usepackage{amssymb}
\usepackage{epsfig}
\usepackage{subfigure}
\usepackage{float}
\usepackage{wrapfig}
\usepackage{color}

\begin{document}
\title{Packing, alignment and flow of shape-anisotropic grains in a 3D silo experiment}
\author{Tam\'as B\"orzs\"onyi,$^{1}$ Ell\'ak Somfai,$^{1}$ Bal\'azs Szab\'o,$^{1}$ Sandra Wegner,$^{2}$ Pascal Mier,$^{3}$  Georg Rose$^{3}$ and Ralf Stannarius,$^{2}$}
\address{
$^1$Institute for Solid State Physics and Optics, Wigner Research Center for Physics, Hungarian Academy of Sciences, P.O. Box 49, H-1525 Budapest, Hungary
$^2$Institute of Experimental Physics, Otto-von-Guericke-University, Universit\"atsplatz 2, D-39106 Magdeburg, Germany,
$^3$Institute of Medical Engineering, Otto-von-Guericke-University, Universit\"atsplatz 2, D-39106 Magdeburg, Germany
}

\ead{borzsonyi.tamas@wigner.mta.hu}
\begin{abstract}
Granular material flowing through bottlenecks like the openings of silos tend to clog and to inhibit further flow.
We study this phenomenon in a three-dimensional hopper for spherical and shape-anisotropic particles by means of X-ray
tomography. The X-ray tomograms provide information on the bulk of the granular filling, and allows to determine the
particle positions and orientations inside the silo. In addition, it allows to calculate local packing densities in
different parts of the container.
We find that in the flowing zone of the silo particles show a preferred orientation and thereby a higher order.
Similarly to simple shear flows, the average orientation of the particles is not parallel to the streamlines but encloses
a certain angle with it. In most parts of the hopper, the angular distribution of the particles did not reach the one
corresponding to stationary shear flow, thus the average orientation angle in the hopper deviates more from the streamlines
than in stationary shear flows.  In the flowing parts of the silo shear induced dilation is observed, which is more pronounced
for elongated grains than for nearly spherical particles.
The clogged state is characterized by a dome, i. e. the geometry of the layer of grains blocking the outflow.
The shape of the dome depends on the particle shape.
\end{abstract}
\newpage

\section{Introduction}

Silo discharge and the flow of granulates through narrow orifices are important ubiquitously in
everyday life and in technological processes. A huge number of experimental, as well as numerical studies have
been performed over many years to understand the underlying processes
(e. g. \cite{Grudzie2011,JIN_2010,mankoc_2009,mankoc_silo,saraf_2011,Tao_2010,Unac_2012,Lastakowski2015,%
gutierrez_2015,Wang201543,Lozano_2012,janda_2009,To_2001,%
aguirre_2014,Wilson2014,Thomas2013,Thomas2015,Tang_Behringer_2011,zuriguel_SR,arevaloSM2016,Lozano2015,Rubio_largo_2015}).

The spontaneous formation of arches~\cite{Tang_Behringer_2011,zuriguel_SR}, the kinetics preceding arch formation
\cite{Rubio_largo_2015}, forces and force chains~\cite{Hidalgo_2013,Vivanco_2012}, and arch breaking \cite{Lozano_2012,janda_2009,Lozano2015} are among the important features when one is interested in the prevention of
silo clogging. The formation of arches, i. e. arrangements of grains at the outlet of a two-dimensional hopper that
block outflow, was analyzed in a pioneering study by To et al.~\cite{To_2001}. For disks in a two dimensional silo,
they found that the arch span is always larger than the opening of the silo and that for frictionless particles the
arch is convex everywhere. Introducing friction they found clogging events where the arch at the opening may not be
convex locally. In addition, they determined the jamming probability as a function of the size of the hopper
opening. Clogging probabilities were also measured not only for spheres but also for
one type of elongated grains by Zuriguel et al.~\cite{zuriguel_2005_silo} in a three dimensional silo.
In their model, the mean avalanche size between two clogging events could be fitted with a power law. The model
assumes a critical radius $R_c$ of the orifice, above which the hopper outflow is continuous, without clogging.
However, the divergence of the mean avalanche size with increasing orifice radius $R$ according to the model
in Ref.~\cite{zuriguel_2005_silo} is proportional to  $(R_c-R)^{-7}$, so it is very difficult to confirm the
existence of such a critical radius experimentally with reasonable amounts of grains available.
Recent work by Thomas and Durian \cite{Thomas2013,Thomas2015} brought new insights in clogging probabilities,
flow rates and mean avalanche sizes in hoppers of different geometries.
They proposed a description of clogging events based on an assumed Poisson distribution of clogging events.
Their model yields an exponential increase of avalanche sizes with increasing hopper orifice radius, so that
no critical $R_c$ exists. These two descriptions are not much distinct from a practical point of view,
when one assumes that the critical hole size is defined by clogging events becoming highly improbable and
practically unobservable above a certain radius.

Since silo discharge is important for many processes in everyday life and the industry, a large number of
studies have been performed both experimentally and numerically over years in order to understand
silo dynamics~\cite{Grudzie2011,JIN_2010,mankoc_2009,mankoc_silo,saraf_2011,Tao_2010,Unac_2012,Lastakowski2015,%
gutierrez_2015,Wang201543}, including horizontal apertures~\cite{aguirre_2014} or tubes~\cite{janda_2015}.
Related phenomena where a distinct number of objects has to pass a bottleneck, for example escape panics~\cite{Helbing_2000},
traffic jams~\cite{kerner_1996} and animal behavior~\cite{Garcimartin_2015,zuriguel_SR}, are in several respects similar to silo
outflow.

The geometrical features of force chains and arches before and in the clogged state can be visualized
using photoelastic disks \cite{Vivanco_2012,Tang_Behringer_2011,zhangSM2014} in 2D hoppers. During the outflow, arches form
in the granular material even in regions well above the orifice of the silo, but they are dynamic and form only
intermittent networks of contact forces. The outflow is not controlled by this network, but the network of
contact forces is responsible for fluctuations in the velocity field. Clogging occurs in 2D
by the formation of an arch with an uninterrupted force chain at the orifice. This structure
remains static unless it is broken by external forces. In 3D, one expects an analogous scenario.
Internal force chains are much more difficult to visualize there than in 2D, therefore a detailed characterization of
3D clogs is still missing. We will refer to these structures in the following as domes, the 3D analogies of a 2D arch,
meaning a shell of grains at the outlet forming a closed net of force chains above the orifice.

Until now, many investigations of hopper flow were focused on spherical (3D) or disk shaped (2D) particles
with only a few studies providing data for shape-anisotropic grains, 
for example Refs.~\cite{Cleary_conf,Cleary_AMM,Li_2004,Liu2014,Langston2004,Kanzaki_2011}. In order to quantify how the discharge
rate is affected by changing the particle aspect ratio, numerical (DEM) methods are the most appropriate.
Here one can tune the aspect ratio by leaving the other important parameters (particle volume, surface friction, etc.)
the same. The aspect ratio dependence is not well understood yet, as in 2D systems Cleary and Sawley
found reduced flow rates for elongated particles with {\em nonzero} friction \cite{Cleary_conf,Cleary_AMM}, while Langston
found increased flow rates for elongated particles with {\em zero} friction \cite{Langston2004} compared to the case of circles.
In 3D simulations Langston found the same flow rate for sphere and sphero-cylinders with zero friction \cite{Langston2004},
while Liu found a reduced flow rates for both prolate and oblate ellipsoids with nonzero friction \cite{Liu2014}.
On the other hand Li reported increased flow rates for oblate ellipsoids with nonzero friction \cite{Li_2004}.
The geometry of the flow field also depended on the aspect ratio, as the flow was rather concentrated to a central
zone for nonspherical grains resulting in a larger stagnant zone near the silo walls \cite{Cleary_conf,Cleary_AMM,Liu2014}.
The consequences of anisometric grain shape for the static and dynamic properties of granular systems have been highlighted in
various recent communications \cite{Tao_2010,JIN_2010,tamas_2012_prl} and reviews \cite{tamas_review,Lu2015}.
Some examples are effects of particle elongation on disordered packing \cite{donev_2004}, self-organized criticality
on a heap \cite{frette_1996}, ordering in shaken systems \cite{narayan_2006} or secondary convection in cylindrical
shear flows \cite{wortel_2015}.

In the present experimental study, we analyze the effect of shape anisometry of granular particles on hopper flow
and clogging by means  X-ray tomography. For this purpose we use elongated pegs, flat lentils and nearly
spherical peas. These particles have sizes of several millimeters, and can be detected and
individually distinguished in the recorded X-ray tomograms. The silo is a cylindrical bucket with changeable orifice size.
The size is chosen such that clogging occurs after finite avalanche sizes. The tomograms are recorded in the clogged
state with the grains at rest.
From comparison of subsequent clogs, we can characterize the flow field
(streamlines) of the granulate in the silo. We find all positions and orientations of individual particles in the
clogged state, and use the tomographic data to calculate the packing fractions and particle orientations. The tomograms
also give information about the shape of the domes, as well as on the orientation of particles forming the dome.

\section{Experiment}

\begin{figure}[htbp]
 \centering
        \includegraphics[width=0.52\columnwidth]{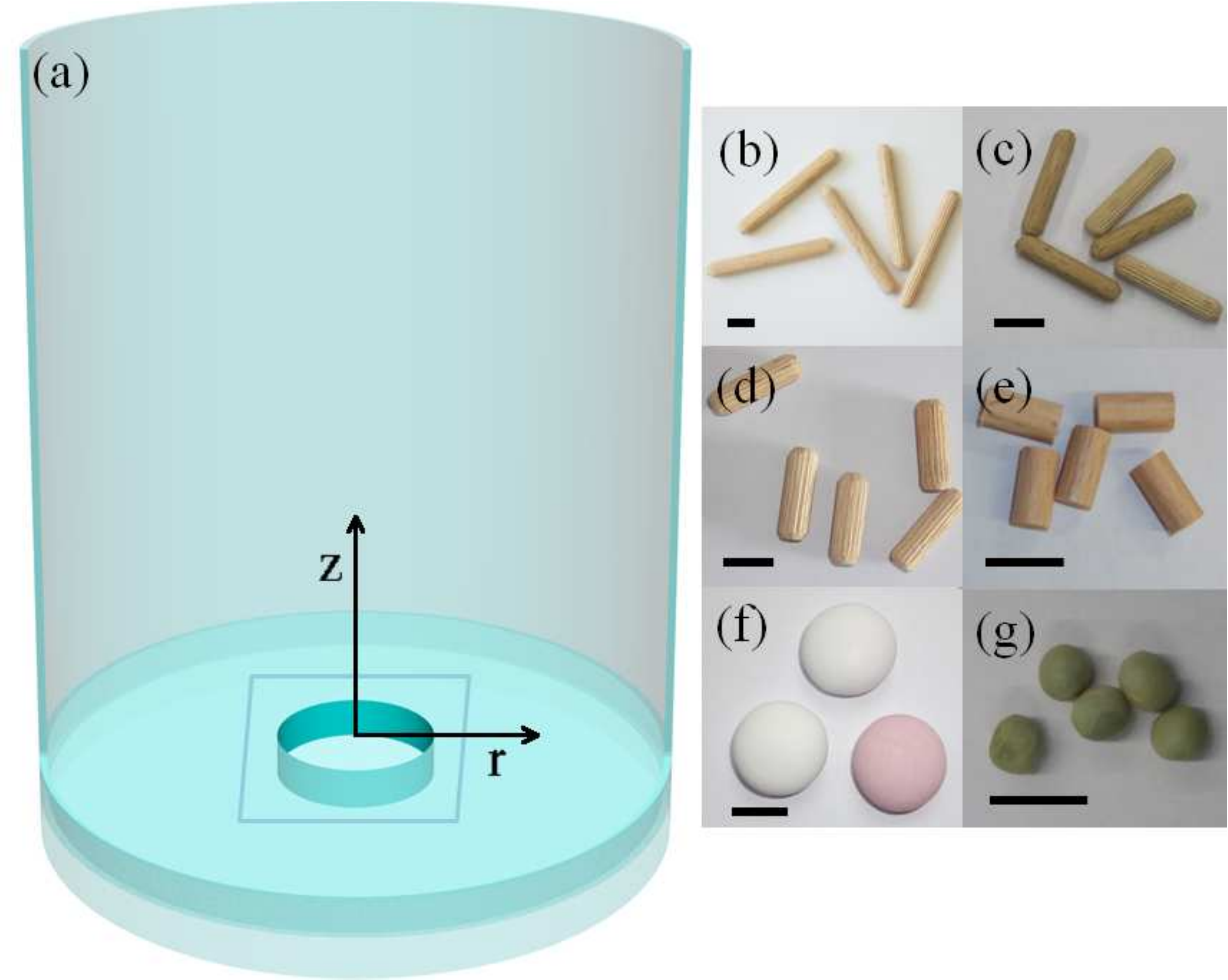}
        \caption{(a) Schematic drawing of the silo, at the bottom the removable plate is sketched.
                 (b-g) Photographs of particles studied, pegs with aspect ratios $Q = \ell/d$
                 (b) $Q=8$, (c) $Q=5$, (d) $Q=3.3$, and (e) $Q=2$, (f) chocolate lentils with $Q\approx 0.45$,
                 (g) peas with $Q\approx 1$.
                 %(h) glass rods  $Q =3.5$ and (i) airsoft munition with $Q=1$, perfect spheres.
                 The horizontal bars correspond to 1 cm.}
        \label{fig:aufb1}
\end{figure}

The experiments were performed with a small plastic silo consisting of a bucket of a diameter of $19$~cm and a height
of $21.4$~cm. The aperture of the outlet was adjustable by exchanging insets in the center of the bottom
plane. The bottom of the bucket has a quadratic opening of about $5 \times 5$~cm$^2$, where different plates with
circular holes were inserted, hole sizes could be chosen in 0.5~mm radius steps. A schematic drawing of the silo is
shown in Fig.~\ref{fig:aufb1}.

%%%%% fig1

The different granulates studied in this setup are also shown in Fig.~\ref{fig:aufb1}. To characterize the deviation
from the spherical shape we use two different quantities: (1) the aspect ratio $Q$, which is the ratio of the size of
the particle along its rotational axis and perpendicular to it and (2) the equivalent radius $r_\mathrm{eq}$ which is
the radius of a sphere with the same volume as the particle. The following materials were used:

\begin{itemize}
\item
wooden pegs with cylinder shape and tapered ends, diameter $d=5$~mm, length $\ell=40$~mm, aspect ratio $Q=\ell/d= 8$, equivalent radius $r_\mathrm{eq}=5.7$~mm
\item
wooden pegs with cylinder shape and tapered ends, $d=5$~mm,  $\ell=25$~mm, $Q=5$, $r_\mathrm{eq}=4.9$~mm
\item
wooden pegs with cylinder shape and tapered ends, $d=6$~mm,  $\ell=20$~mm, $Q\approx 3.3$, $r_\mathrm{eq}=5.1$~mm
\item
wooden pegs with cylinder shape,   $d=5$~mm,  $\ell=10$~mm,  $Q=2$, $r_\mathrm{eq}=3.6$~mm
\item
ellipsoidal chocolate lentils ({\em Piasten}) covered with hard icing with $d=18.5$~mm and $h=8.3$~mm, $Q=0.45$, $r_\mathrm{eq}=7.1$~mm
\item
peas, with small deviations from a perfect sphere, polydisperse with mean diameter of 7.6~mm, standard deviation 0.23~mm, $r_\mathrm{eq}= 3.8$~mm
\end{itemize}

Two types of pegs (Q = 8, 3.3) and some of the particles of the other two types (Q =5, 2) have slight axial groves
on their surfaces.

For the preparation of the experiment, the silo outlet is closed first. Then the container is filled with the grains.
We have selected the outlet sizes for the individual grain types such that after the bottom hole is opened, the
granulate flows out but clogs after an appropriate time. The conditions for a well prepared state are that the avalanche
is sufficiently large to create a representative flow and alignment pattern in the container, and that
the avalanche stops by clogging when there is still enough material in the container cover more than the region  covered
in the subsequent tomogram. In this clogged state, an X-ray tomogram of the lower half of the silo is recorded.
We use the robot-based flat panel X-ray C-arm system Siemens Artis zeego of the STIMULATE-lab, Otto von Guericke University,
Magdeburg. The chosen spatial resolution was 2.03 pixel/mm, with recorded volumes
of 25.2 cm $\times$ 25.2 cm $\times$ 19 cm. The 3D arrangement of the particles is then determined from the X-ray tomogram.

To calculate the packing density in different parts of the silo, the tomograms are binarized. We define a threshold for
all voxels, that determines whether the voxel belongs to a grain or not. For that we have to choose carefully the threshold.
Our algorithm is based on Otsu's method~\cite{Otsu1979} and to find a good threshold, we only take the part of the tomogram
with particles in it. Borders and empty regions are avoided. Afterwards the processed tomograms are averaged for each
material. After this the 3D image is projected to a 2D representation where we average over all equivalent voxels,
i. e. voxels with the same distance to the center of rotation and the same height. This method gives a measure of the relative packing densities in different parts of the silo.

\begin{figure}[b]
\centering
\includegraphics[height=4cm,clip]{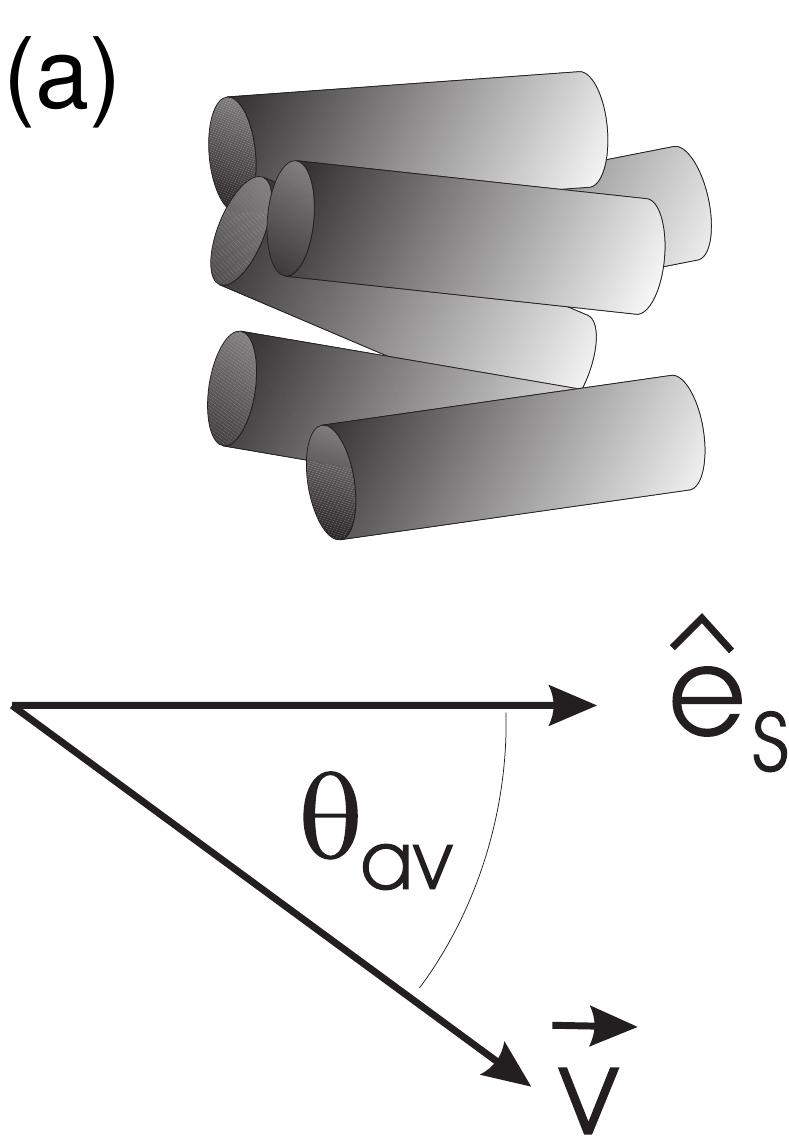}\hspace{2cm}
\includegraphics[height=4cm,clip]{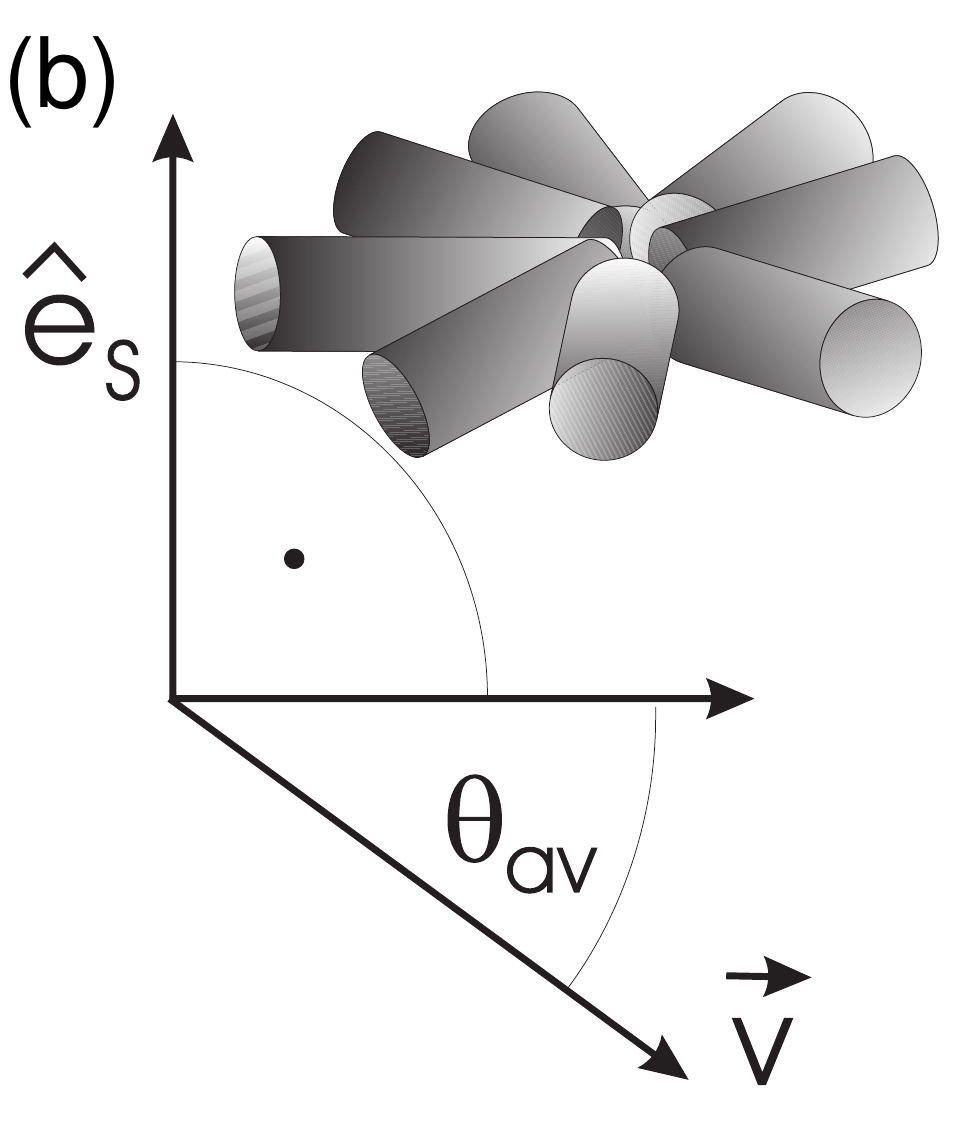}
\caption{Diagram showing the definition of the flow alignment angle $\theta_{\rm av}$
from the principle axis $\hat e_S$ of $S$ and the flow direction $\vec v$ for (a) positive order parameter $S>0$,
and (b) negative  order parameter $S<0$.
}
\label{fig:diagram}
\end{figure}

In order to calculate the streamlines, the nematic order parameter and the
average orientation angle, we determined the position and the orientation of
the individual particles; for details see \cite{wegner_SM_2014}.
The shear-induced orientational order is monitored by diagonalizing the
symmetric traceless nematic order tensor $\boldsymbol{T}$:
\begin{equation}
T_{ij}= \frac{3}{2N} \sum\limits_{n=1}^N \left[{\ell}^{(n)}_i {\ell}^{(n)}_j -\frac{1}{3} \delta_{ij}
\right] \quad ,
\end{equation}
where $\vec {\ell}^{(n)}$ is the unit vector along the long axis of particle
$n$, and the sum is over all $N$ detected particles in equivalent regions of
the silo. For that purpose, summation over a large number of particles is
needed. As a compromise, we make the assumption that the flow in the axially
symmetric geometry of our silo yields an axially symmetric mean flow field and
consequently an axially symmetric mean arrangement of the particles, therefore,
we average over ring-shaped zones of the silo.

Conventionally, the nematic order parameter $S$ is the eigenvalue of the order
tensor with the largest absolute value. 
In the case $S>0$ (preferential alignment along a certain axis), we will refer to
$S$ in the definition of flow alignment and define the average flow orientation angle 
$\theta_{\rm av}$ as the angle between $\hat e_S$ (the eigenvector of $\boldsymbol{T}$ 
corresponding to $S$) and the local flow direction, see Fig.~\ref{fig:diagram}(a).

In regions with strong planar influences, i.e. near walls and close to
the free surface, the orientational ordering of the particles is often fan-like:
the axes of the particles are preferentially aligned perpendicular to a certain axis,
but azimuthally more or less distributed at random.
In that case, $\boldsymbol{T}$ has two small positive eigenvalues, and the third one ($S$) is negative and has
the largest absolute value (recall that $\boldsymbol{T}$ is traceless). In that case, we use a different, more
descriptive concept to characterize the local alignment:
The eigenvector $\hat e_{S}$ corresponding to $S$ marks the direction in which the particles are preferentially \emph{not} oriented. Therefore, $\theta_{\rm av}$ is defined differently, as the angle between the
streamlines and the closest vector in the orientational plane (plane perpendicular to $\hat e_{S}$),
see Fig.~\ref{fig:diagram}(b). In this case, we characterize the flow-induced order by the largest positive eigenvalue $S'$.
This helps us to visualize shear induced orientational changes in the entire hopper, since $S'$ is less susceptible to boundary 
effects.

\section{Results and Discussion}

\subsection{Order, Orientation and Packing Fraction for Pegs}

\begin{figure}[htbp]
 \centering
        \includegraphics[width=0.6\textwidth]{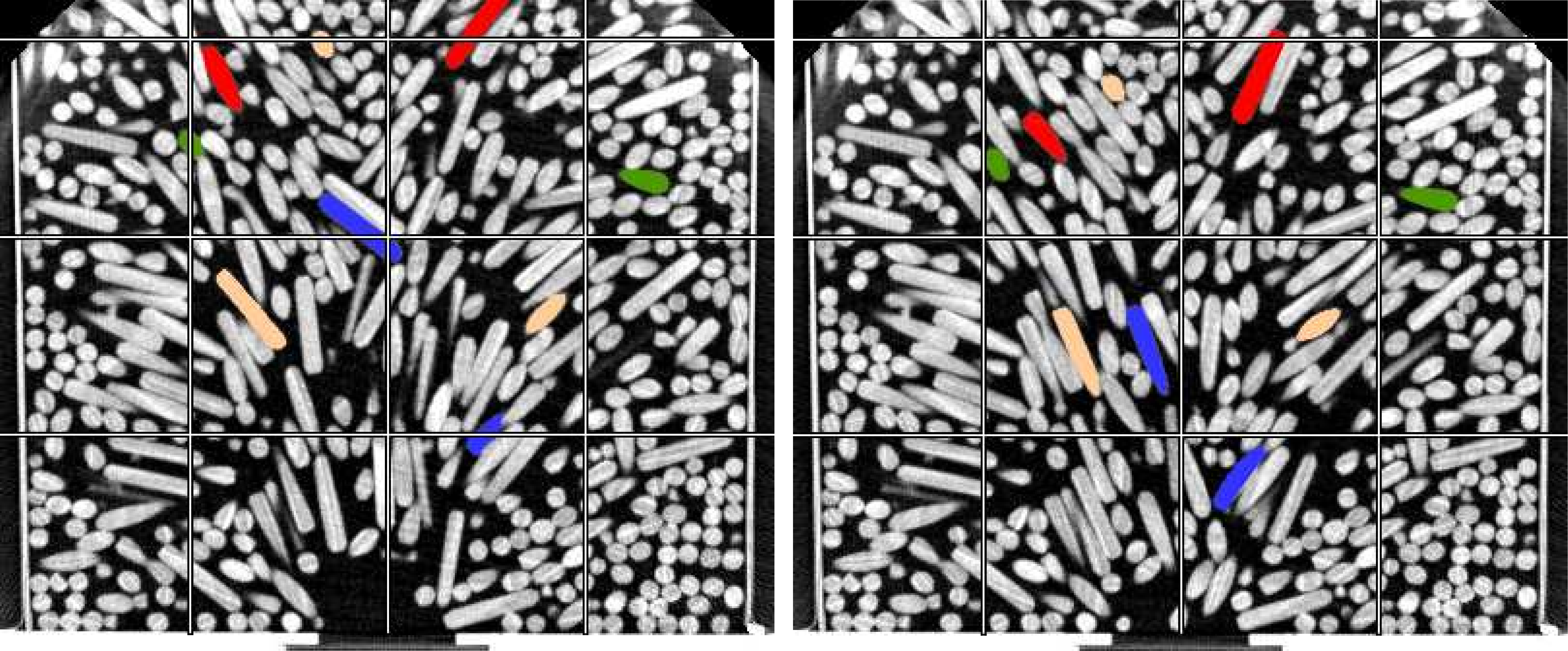}
        \caption{Central vertical cuts of two subsequent tomograms (two clogged states separated by a single avalanche) for
                 pegs with aspect ratio $Q=5$.  In the flowing region particles are aligned. The pictures show the lower 16 cm of the
                 nearly cylindrical hopper, a rectangular mesh of 5 cm $\times$ 5 cm  has been superimposed to guide the eye.
                 A few colored grains indicate the displacement of selected particles between the two subsequent clogged states.
                 The diameter of the hopper outlet was 35 mm.}
        \label{fig:bsp_q5_tomogramm}
\end{figure}

Central vertical cuts of two subsequent tomograms are presented for pegs with aspect ratio $Q=5$ in
Fig.~\ref{fig:bsp_q5_tomogramm}.
%%%%% fig3
These tomograms were taken after several avalanches, thus the orientation and packing
%%%%% fig4
\begin{figure*}[htbp]
        \includegraphics[width=\textwidth]{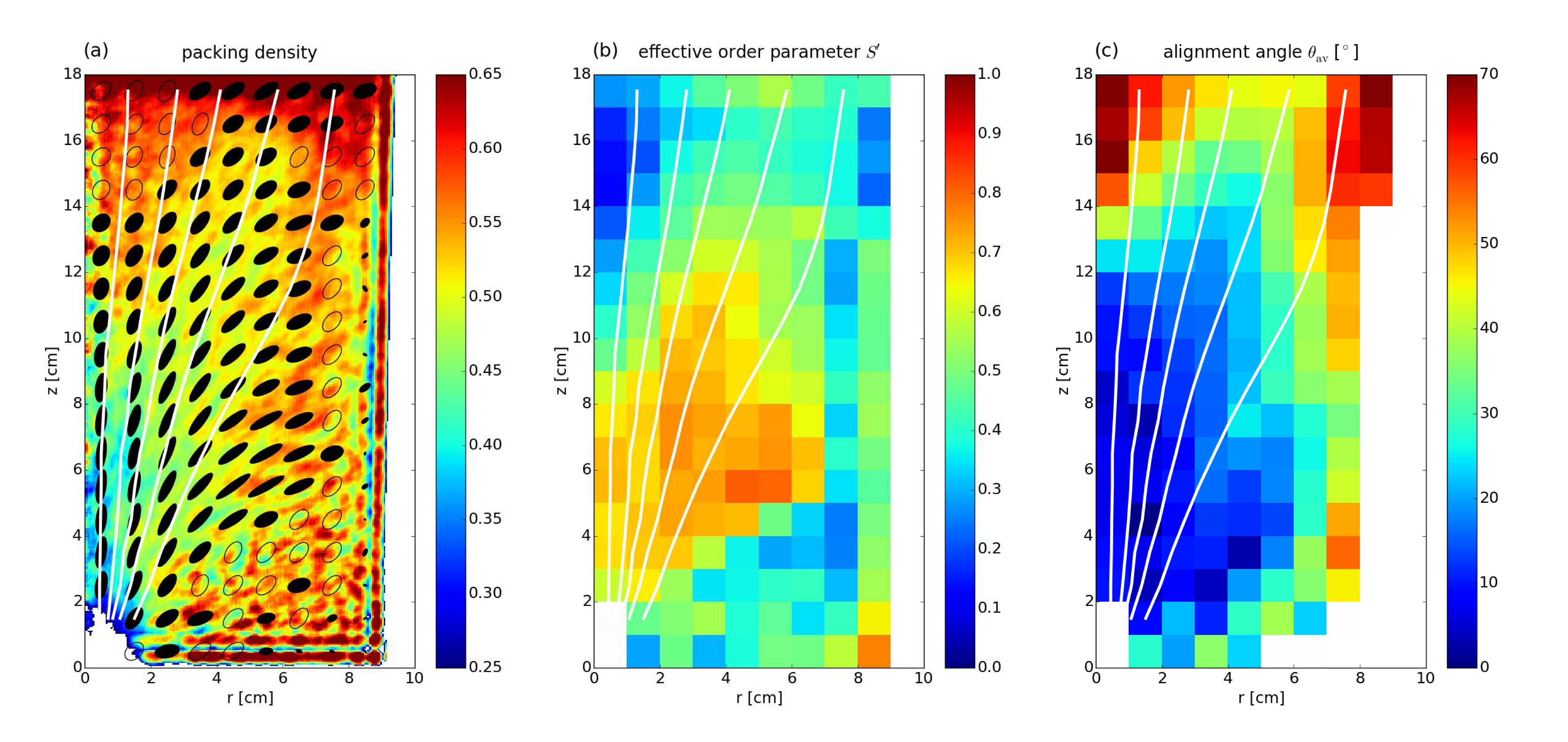}
        \caption{Color maps: distributions of (a) packing density, (b) effective order parameter $S'$ and (c) average alignment angle
                 $\theta_{\rm av}$ in the $r-z$ plane. The alignment angle is measured with respect to the streamlines indicated
                 by white lines. On panel (a), the alignment and ordering of the particles is also represented by ellipses (see text).
                 Data have been calculated from 26 subsequent tomograms (examples shown in Fig.~\ref{fig:bsp_q5_tomogramm}) for
                 pegs with $Q=5$ and hopper outlet of 35 mm.
                 Correlation between flow and orientation of the grains is given in Fig.~\ref{fig:theta-streamlines}
                 }
        \label{fig:theta_s_q5}
\end{figure*}
of the grains reflects the properties of the developed flow (initial conditions erased in the flowing regions).
As explained above, these tomograms were taken in the clogged state. During the avalanche between these two
tomograms, approximately 60 particles left the hopper.
The displacement of the material due to this avalanche is visualized by a few selected particles which are marked
with colors in Fig.~\ref{fig:bsp_q5_tomogramm}. It is clearly visible that the central part moves faster, especially
approaching the outlet. Stagnant zones can be identified near the walls especially at the bottom of the hopper.
In the flowing region particles are expected to be oriented nearly (but not exactly) parallel to the streamlines
\cite{tamas_2012_prl,wegner_2012,tamas_2012_pre}.

In order to quantify the ordering and alignment of the particles and determine the flow field and the density distribution,
tomograms from 26 subsequent clogged states have been analyzed. In Figures \ref{fig:theta_s_q5}a-c, the colormaps indicate
the density distribution, the effective order parameter ($S'$) and the average alignment angle ($\theta_{\rm av}$) of 
the particles in
the $r-z$ plane. The alignment angle is defined in the interval $-90^\circ<\theta<90^\circ$.
The alignment angle is measured with respect to the  streamlines, which are indicated by white lines.
In order to better visualize the correlation between these quantities, on Fig.~\ref{fig:theta_s_q5}a we represented the two
other parameters (the average orientation of the particles and the nematic order parameter) by ellipses.
The size and orientation of the ellipse represent $S$ and $\theta_{\rm av}$ in the following way.
The area of the ellipse is proportional to the projection of the average orientation of the particles to the $r-z$ plane.
In the flowing region the rods are aligned due to shear in this plane, so we find ellipses with large area.
Near the vertical walls, however, the particles are mostly oriented tangentially, thus the area of the ellipse is small.
The direction of alignment is reflected in the orientation of the ellipses.
The order parameter is represented by the flattening of the ellipses: $S =$ flattening $= 1 - b/a$,
where ``$a$'' and ``$b$'' are the lengths of the long and short semi-axes, respectively.
Thus $S=1$ (perfect order) would be represented by a thin line, and
$S=0$ (isotropic state) by a circle. 
All the above applies for usual (calamitic) nematic order,
$S >0$, where the largest absolute-value eigenvalue of $T$ coincides with the largest eigenvalue (full ellipses).
For the fan-shape oriented case, where the eigenvalue with largest absolute-value is
negative (${S}<0$), we draw an empty ellipse with flattening $|S|  = 1 - b/a$.
In those cases the short (of length ``$b$'') semi-axis is oriented to the axis of the corresponding eigenvector of
the order tensor.
At these locations, the history of the filling procedure or boundary effects produced an initial 2D order,
where the orientation of the particles are nearly within a plane, resulting in a negative order parameter.
To quantify the flow induced ordering it is more suitable to take the largest positive eigenvalue as an effective order 
parameter $S'$.

Altogether, Figure \ref{fig:theta_s_q5} shows that in the flowing (sheared) regions we find well oriented rods due to
the shear flow, while near the walls tangential alignment is preferred due to the boundary conditions.
We see, that in the shear flow the average orientation is not parallel to the streamlines,
but there is a flow alignment angle as expected \cite{tamas_2012_prl,wegner_2012,tamas_2012_pre}.
The sheared regions are characterized with lower packing fraction, consistent with our earlier observation
in shear flow, where the shear induced dilation was found to be only partially compensated with
a slight density increase due to the ordering of the grains \cite{wegner_2012}.

In order to accurately characterize the alignment of the particles in this complex flow field, let us recall earlier
 %%%%% fig5
\begin{figure}[htb!]
 \centering
        \centerline{\includegraphics[width=0.9\columnwidth]{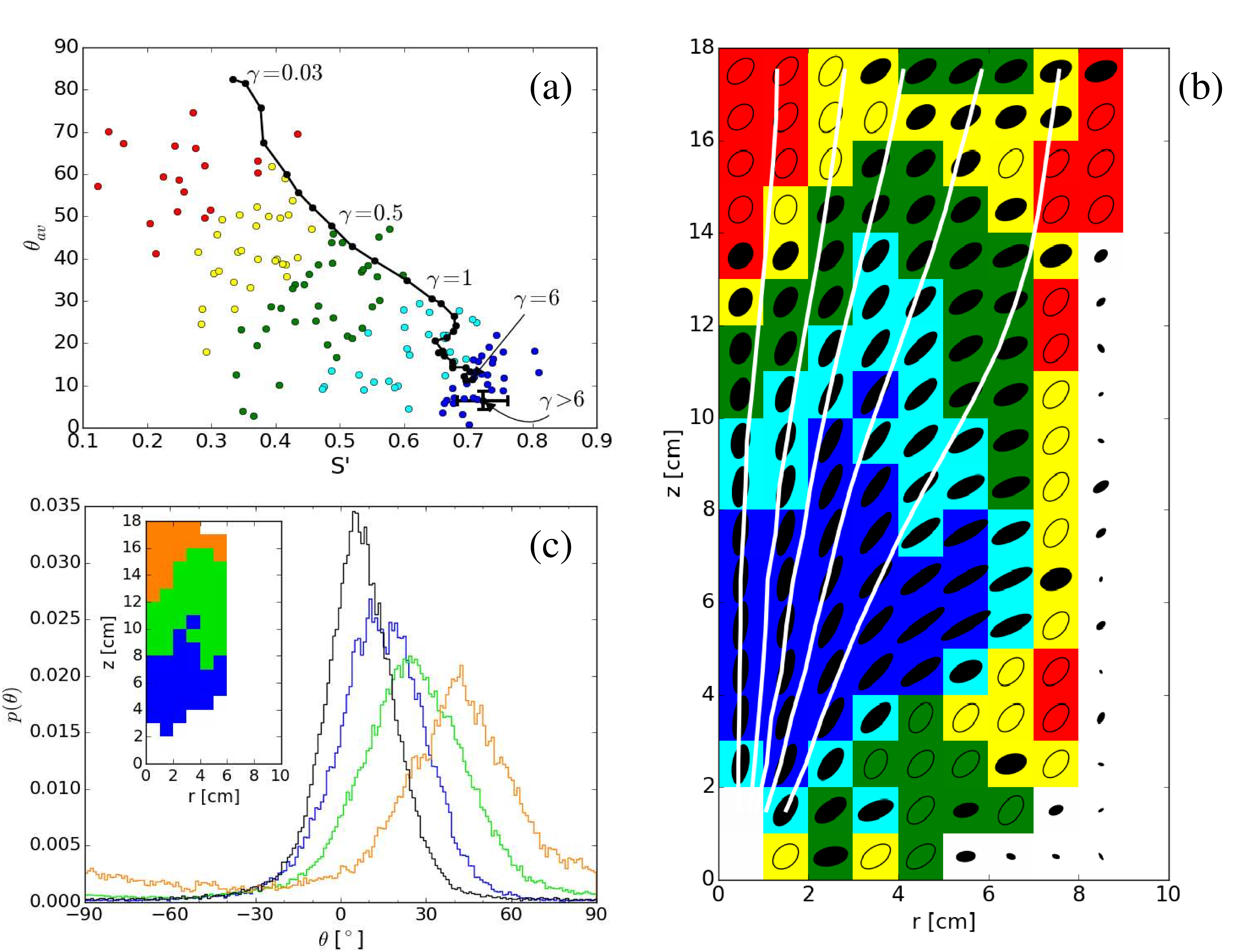}}
        \caption{(a) Average particle orientation $\theta_{\rm av}$ as a function of the order parameter $S'$. Data from different parts
                     of the hopper are marked with different colors, these regions are marked with the same color in panel (b).
                     The black line in panel (a) represents the data from the split bottom Couette flow. In panel (b) $S$ and
                     $\theta_{\rm av}$ are marked with ellipses and the streamlines with white lines as in Fig.~\ref{fig:theta_s_q5}.
                     (c) Distribution of particle orientations $\theta$ with respect to the streamlines. The inset shows the 3 regions in the
                 hopper corresponding to the 3 curves with different colors. The black curve corresponds to the stationary distribution
                 of the same particles in split bottom Couette flow \cite{tamas_2012_pre}.
                     }
        \label{fig:theta-streamlines}
\end{figure}
data on the behaviour of these particles in a split bottom Couette flow \cite{szabo_2014_pre}. In that experiment, we
sheared the same material in a cylindrical geometry and recorded how the orientational order develops from an initially random sample.
This dataset is presented in Fig.~\ref{fig:theta-streamlines}a by a black line. As the local shear deformation $\gamma$ is increasing,
the order parameter rapidly increases and by $\gamma=1$, it already reaches about $S=0.6$. During this time, the average alignment
angle strongly decreases from about $\theta_{\rm av}=80^{\circ}$ to about  $\theta_{\rm av}=35^{\circ}$ as a consequence of the
continuous rotation of the rods due to the shear flow. When the sample is sheared further (above $\gamma=6$), a stationary state
is reached which is characterized by a shear alignment angle of about $\theta_{\rm av}=8^{\circ}$. In this state
the particles still rotate, but their rotational velocity strongly depends on their orientation, with slow rotation around $\theta=8^{\circ}$, resulting in an ensemble averaged alignment of the sample.
This dataset was obtained from 90 X-ray tomograms in the cylindrical split bottom geometry.

Focusing on the orientation of the particles in the hopper (see colored data points in Fig.~\ref{fig:theta-streamlines}a),
we find that a good part of the sample is well ordered with similar values of $S$ and $\theta_{\rm av}$ (see blue
data points) as in stationary shear flow. This is the region right above the outlet (indicated with blue) in
Fig.~\ref{fig:theta-streamlines}b, where the sample was exposed to the largest shear deformation. Further upstream where the
material was sheared less, we find a smaller order parameter and a larger alignment angle, similarly to the initial stages of the
case of simple shear (see regions with light blue, green, yellow and red colors in Fig.~\ref{fig:theta-streamlines}a-b).
In order to better track the evolution of the particle orientation in the hopper flow,
we present the histogram of the
particle orientations for 3 regions in our hopper in Fig.~\ref{fig:theta-streamlines}c.
The three regions are marked with
%%%%% fig6
\begin{figure*}[htb!]
\includegraphics[width=\textwidth]{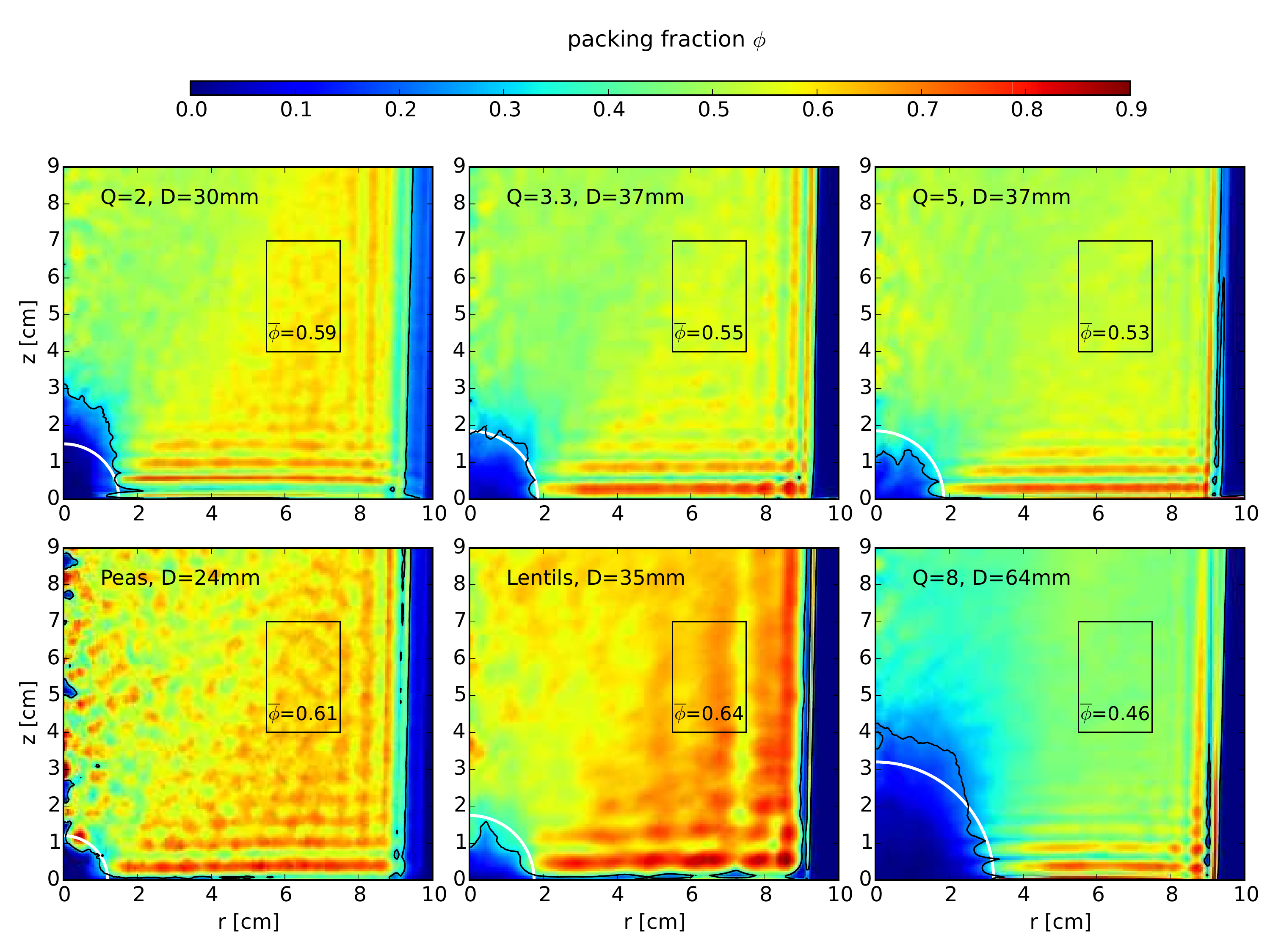}
\caption{Distributions of the packing fraction in the $r-z$ plane for elongated particles: pegs with aspect ratios $Q=2$, $Q=3.3$, $Q=5$ and
$Q=8$, as well as peas and lentils.
    The diameter of the hopper opening is indicated on the top of each image.
    The density in the stagnant zone $\overline{\phi}$ was defined by averaging in a rectangular box as indicated on the images.
        The black level set drawn at $0.5 \overline{\phi}$ indicates the shape of the dome. A hemisphere of equal diameter with the outlet
        is indicated by a white line for comparison.  Each image represents an average of at least 30 tomograms.
\label{fig:dichten_alle}}
\end{figure*}
different colors (see the inset of Fig.~\ref{fig:theta-streamlines}c). The corresponding 3 curves show gradual narrowing of the
angular distribution and decreasing of the average angle as the material is exposed to larger and larger deformation
as it is displaced downwards in the hopper. In Fig.~\ref{fig:theta-streamlines}c we also show the histogram of angles
of stationary split bottom Couette flow (black line). We see that near the outlet (blue curve) the angular distribution
almost reached the case of stationary shear.
Coming back to Fig.~\ref{fig:theta-streamlines}a, it is remarkable, that the data points taken in the hopper are almost all below
the black line (corresponding to the case of simple shear).
Physically, the difference between the two cases is, that in the hopper the streamlines are strongly converging while
in the cylindrical split bottom geometry they run parallel with each other. Thus in the hopper the converging
flow leads to particle orientations which are closer to the streamlines than in a simple shear flow.

\subsection{Packing Fraction for Various Materials}

So far we have presented the flow and orientation fields and the packing density of the particles for one type of pegs ($Q=5$)
in a hopper.  For this we used 26 tomograms, with small particle displacement (small avalanches) between the tomograms,
so that the flow lines could be identified.
In the following, we present the packing density distribution in a hopper for six materials.
For these plots, a similar number of tomograms was taken for each material, but here we allowed more grains to flow out
of the hopper between subsequent measurements (compare outlet diameters) to reduce
correlation in the arrangement of the grains. The resulting packing density plots are shown in Fig.~\ref{fig:dichten_alle} for pegs
with aspect ratio $Q=2$, $Q=3.3$, $Q=5$ and $Q=8$, for lentils and peas.

As it is seen in  Fig.~\ref{fig:dichten_alle}, the density is reduced in the sheared region (compared to the density in the
stagnant zone near the walls) for all materials due to the shear induced dilation.
As expected, the density in the stagnant zone $\overline{\phi}$ decreases with increasing aspect ratio for all rodlike grains.
The region where $\overline{\phi}$ was determined is marked with a black rectangle on the images.
The shear induced dilation is stronger for rods, compared to the case of nearly spherical particles (peas).
Approaching the outlet from the top along the symmetry axis of the hopper we find a decreasing packing density.
Near the outlet there is a practically empty region (marked with blue), the region below the dome. The upper border
of this region corresponds to the dome which holds the material above in this clogged configuration.
We indicated this with a black line, corresponding to $50\%$ of the packing fraction observed in the stagnant zone.
The contour of a hemisphere of equal diameter with the outlet is also marked (as a white line) to visualize deviations from a spherical shape.
Analyzing the shape of the dome, we find that its height decreases with particle elongation for the  sequence $Q=2 \to 3.3 \to 5$.
For lentils and nearly spherical peas the dome is similar to the case of $Q=3.3$.

For pegs with $Q=8$ we find a higher dome, we note however, that these particles are relatively large compared to the system size,
thus boundary effects might already have an influence in this case.
We should also note that boundaries also induce layering of the particles especially near the bottom wall, where anisometric
particles lay flat.
In the stagnant zones near the boundaries the inhomogeneities of the density field are stronger compared to the flowing regions,
since there is less averaging over different configurations.

%---------------------------------------------------------------------------------------------------------
\section{Summary}
We presented experimental data about the packing fraction, grain alignment, orientational order parameter, and flow field in
a 3D hopper based on X-ray CT measurements. We analyzed subsequent clogged states for 6 materials including elongated particles
(pegs), lentils, and nearly spherical grains (peas). We have shown that for elongated particles the grains get ordered in the
flowing parts of the silo. Similarly to the case of simple shear flows the average orientation of the rods is not parallel to the
streamlines but encloses a small angle with it.
The order parameter increases as the grains travel downwards the silo and the local shear deformation grows. In most parts of the hopper
the orientational distribution of the grains did not reach the stationary orientational distribution observed for simpler shear flows.
Consequently, in these regions the order parameter is smaller and the average alignment encloses a larger angle with the streamlines than in
a simple shear flow.
The packing density of the material is smaller in the flowing (sheared) regions compared to the stagnant zones.
This density decrease is stronger for elongated grains compared to the case of nearly spherical particles (peas).
Along the vertical symmetry axis of the hopper the density continuously decreases from top to bottom.

Near the outlet the density distribution averaged over many clogged configurations provides information about the shape of a typical dome,
i.e. the last layer of the grains which blocked the flow. A characteristic feature is the ratio of hole diameter and elevation
of the dome above the outlet in the center. We find that the height of the dome decreases with increasing aspect ratio for pegs with
$Q=2$ $\rightarrow$ $Q=3.3$ $\rightarrow$ $Q=5$. This has consequences for the average number of grains that are involved in the
formation of the clog, and therefore also for the clogging probability, the avalanche statistics and the critical radius for the
divergence of the mean avalanche sizes. The analysis of these details requires not only the information on the dome shape, but
also on the characteristic alignment of particles forming the clog. This will be treated in a subsequent investigation.

%---------------------------------------------------------------------------------------------------------
\section*{Acknowledgments}

Financial support from the DAAD/M\"OB researcher exchange program (Grant
No. 64975), the Hungarian Scientific Research Fund (Grant No. OTKA NN 107737)
and the J\'anos Bolyai Research Scholarship of the Hungarian Academy of Sciences are acknowledged.


\section*{References}

\begin{thebibliography}{10}

\bibitem{Grudzie2011}
K.~Grudzien, M.~Niedostatkiewicz, J.~Adrien, J.~Tejchman, and E.~Maire.
\newblock Quantitative estimation of volume changes of granular materials
  during silo flow using x-ray tomography.
\newblock {\em Chemical Engineering and Processing: Process Intensification},
  50(1):59 -- 67, 2011.

\bibitem{JIN_2010}
B.~Jin, H.~Tao, and W.~Zhong.
\newblock Flow behaviors of non-spherical granules in rectangular hopper.
\newblock {\em Chinese Journal of Chemical Engineering}, 18(6):931 -- 939,
  2010.

\bibitem{mankoc_2009}
C.~Mankoc, A.~Garcimart\'{i}n, I.~Zuriguel, D.~Maza, and L.~A. Pugnaloni.
\newblock Role of vibrations in the jamming and unjamming of grains discharging
  from a silo.
\newblock {\em Phys. Rev. E}, 80:011309, 2009.

\bibitem{mankoc_silo}
C.~Mankoc, A.~Janda, R.~Ar\'evalo, J.~M. Pastor, I.~Zuriguel,
  A.~Garcimart\'{i}n, and Maza D.
\newblock The flow rate of granular materials through an orifice.
\newblock {\em Granular Matter}, 9:407--414, 2007.

\bibitem{saraf_2011}
S.~Saraf and S.~V. Franklin.
\newblock Power-law flow statistics in anisometric (wedge) hoppers.
\newblock {\em Phys. Rev. E}, 83:030301, 2011.

\bibitem{Tao_2010}
H.~Tao, B.~Jin, W.~Zhong, X.~Wang, B.~Ren, Y.~Zhang, and R.~Xiao.
\newblock Discrete element method modeling of non-spherical granular flow in
  rectangular hopper.
\newblock {\em Chemical Engineering and Processing: Process Intensification},
  49(2):151 -- 158, 2010.

\bibitem{Unac_2012}
R.~O. Unac, A.~M. Vidales, and L.~A. Pugnaloni.
\newblock The effect of the packing fraction on the jamming of granular flow
  through small apertures.
\newblock {\em Journal of Statistical Mechanics: Theory and Experiment},
  2012:P04008, 2012.

\bibitem{Lastakowski2015}
H.~Lastakowski, J.-C. G\'eminard, and V.~Vidal.
\newblock Granular friction: Triggering large events with small vibrations.
\newblock {\em Scientific Reports}, 5:13455, 2015.

\bibitem{gutierrez_2015}
G.~Guti\'errez, C.~Colonnello, P.~Boltenhagen, J.~R. Darias, R.~Peralta-Fabi,
  F.~Brau, and E.~Cl\'ement.
\newblock Silo collapse under granular discharge.
\newblock {\em Phys. Rev. Lett.}, 114:018001, 2015.

\bibitem{Wang201543}
Y.~Wang, Y.~Lu, and J.~Y. Ooi.
\newblock A numerical study of wall pressure and granular flow in a
  flat-bottomed silo.
\newblock {\em Powder Technology}, 282:43 -- 54, 2015.

\bibitem{Lozano_2012}
C.~Lozano, G.~Lumay, I.~Zuriguel, R.~C. Hidalgo, and A.~Garcimart\'{i}n.
\newblock Breaking arches with vibrations: The role of defects.
\newblock {\em Phys. Rev. Lett.}, 109:068001, 2012.

\bibitem{janda_2009}
A.~Janda, D.~Maza, A.~Garcimart\'in, E.~Kolb, J.~Lanuza, and E.~Cl\'ement.
\newblock Unjamming a granular hopper by vibration.
\newblock {\em EPL}, 87:24002, 2009.

\bibitem{Lozano2015}
C.~Lozano, I.~Zuriguel, and A.~Garcimart\'{i}n.
\newblock Stability of clogging arches in a silo submitted to vertical
  vibrations.
\newblock {\em Phys. Rev. E}, 91:062203, 2015.

\bibitem{To_2001}
K.~To, P-Y. Lai, and H.~K. Pak.
\newblock Jamming of granular flow in a two-dimensional hopper.
\newblock {\em Phys. Rev. Lett.}, 86:71--74, 2001.

\bibitem{aguirre_2014}
M.~A. Aguirre, R.~De~Schant, and J.-C. G\'eminard.
\newblock Granular flow through an aperture: Influence of the packing fraction.
\newblock {\em Phys. Rev. E}, 90:012203, 2014.

\bibitem{Wilson2014}
T.~J. Wilson, C.~R. Pfeifer, N.~Meysingier, and D.~J. Durian.
\newblock Granular discharge rate for submerged hoppers.
\newblock {\em Papers in Physics}, 6:060009, 2014.

\bibitem{Thomas2013}
C.~C. Thomas and D.~J. Durian.
\newblock Geometry dependence of the clogging transition in tilted hoppers.
\newblock {\em Phys. Rev. E}, 87:052201, 2013.

\bibitem{Thomas2015}
C.~C. Thomas and D.~J. Durian.
\newblock Fraction of clogging configurations sampled by granular hopper flow.
\newblock {\em Phys. Rev. Lett.}, 114:178001, 2015.

\bibitem{Tang_Behringer_2011}
J.~Tang and R.P. Behringer.
\newblock How granular materials jam in a hopper.
\newblock {\em Chaos}, 21:041107, 2011.

\bibitem{zuriguel_SR}
I.~Zuriguel, D.~R. Parisi, R.~C. Hidalgo, C.~Lozano, A.~Janda, P.~A. Gago,
  J.~P. Peralta, L.~M. Ferrer, L.~A. Pugnaloni, E.~Cl\'ement, D.~Maza,
  I.~Pagonabarraga, and A.~Garcimart\'{i}n.
\newblock Clogging transition of many-particle systems flowing through
  bottlenecks.
\newblock {\em Scientific Reports}, 4:7324, 2014.

\bibitem{arevaloSM2016}
R.~Ar\'evalo and I.~Zuriguel.
\newblock Clogging of granular materials in silos: effect of gravity and outlet
  size.
\newblock {\em Soft Matter}, 12:123, 2016.



\bibitem{Rubio_largo_2015}
S.~M. Rubio-Largo, A.~Janda, D.~Maza, I.~Zuriguel, and R.~C. Hidalgo.
\newblock Disentangling the free-fall arch paradox in silo discharge.
\newblock {\em Phys. Rev. Lett.}, 114:238002, 2015.

\bibitem{Hidalgo_2013}
R.~C. Hidalgo, C.~Lozano, I.~Zuriguel, and A.~Garcimart\'in.
\newblock Force analysis of clogging arches in a silo.
\newblock {\em Granular Matter}, 15(6):841--848, 2013.

\bibitem{Vivanco_2012}
F.~Vivanco, S.~Rica, and F.~Melo.
\newblock Dynamical arching in a two dimensional granular flow.
\newblock {\em Granular Matter}, 14(5):563--576, 2012.

\bibitem{zuriguel_2005_silo}
I.~Zuriguel, A.~Garcimart\'{i}n, D.~Maza, L.~A. Pugnaloni, and J.~M. Pastor.
\newblock Jamming during the discharge of granular matter from a silo.
\newblock {\em Phys. Rev. E}, 71:051303, 2005.


\bibitem{janda_2015}
A.~Janda, I.~Zuriguel, A.~Garcimart\'in, and D.~Maza.
\newblock Clogging of granular materials in narrow vertical pipes discharged at
  constant velocity.
\newblock {\em Granular Matter}, 17:545--551, 2015.

\bibitem{Helbing_2000}
D.~Helbing, I.~Farkas, and T.~Vicsek.
\newblock Simulating dynamical features of escape panic.
\newblock {\em Nature}, 407:487--490, 2000.

\bibitem{kerner_1996}
B.~S. Kerner and H.~Rehborn.
\newblock Experimental properties of complexity in traffic flow.
\newblock {\em Phys. Rev. E}, 53:R4275--R4278, 1996.

\bibitem{Garcimartin_2015}
A.~Garcimart\'{\i}n, J.~M. Pastor, L.~M. Ferrer, J.~J. Ramos,
  C.~Mart\'{\i}n-G\'omez, and I.~Zuriguel.
\newblock Flow and clogging of a sheep herd passing through a bottleneck.
\newblock {\em Phys. Rev. E}, 91:022808, 2015.

\bibitem{zhangSM2014}
L.~Zhang, S.~Cai, Z.~Hu, and J.~Zhang.
\newblock A comparison between bridges and force-chains in photoelastic disk
  packing.
\newblock {\em Soft Matter}, 10:109, 2014.

\bibitem{Cleary_conf}
P.~W. Cleary.
\newblock The effect of particle shape on hopper discharge.
\newblock {\em Second International Conference on CFD in the Minerals and
  Process Industries}, pages 71--76, 1999.

\bibitem{Cleary_AMM}
P.~W. Cleary and M.~L. Sawley.
\newblock DEM modelling of industrial granular flows: 3d case studies and
  the effect of particle shape on hopper discharge.
\newblock {\em Applied Mathematical Modelling}, 26(2):89 -- 111, 2002.

\bibitem{Liu2014}
S.~D. Liu, Z.~Y. Zhou, R.~P. Zou, D.~Pinson, and A.B. Yu.
\newblock Flow characteristics and discharge rate of ellipsoidal particles in a
  flat bottom hopper.
\newblock {\em Powder Technology}, 253:70--79, 2014.

\bibitem{Li_2004}
J.~Li, P.~A. Langston, C.~Webb, and T.~Dyakowski.
\newblock Flow of sphero-disc particles in rectangular hoppers- a DEM and
  experimental comparison in 3d.
\newblock {\em Chemical Engineering Science}, 59(24):5917 -- 5929, 2004.

\bibitem{Langston2004}
P.~A. Langston, M.~A. Al-Awamleh, F.~Y. Fraige, and B.~N. Asmar.
\newblock Distinct element modelling of non-spherical frictionless particle
  flow.
\newblock {\em Chem. Eng. Sci.}, 59:425--435, 2004.


\bibitem{Kanzaki_2011}
T.~Kanzaki,  M.~Acevedo, I.~Zuriguel, I.~Pagonabarraga,  D.~Maza, and R.C.~Hidalgo.
\newblock Stress distribution of faceted particles in a silo after its partial discharge.
\newblock {\em Eur. Phys. J. E}, 34:1, 2011.


\bibitem{tamas_2012_prl}
T.~B\"orzs\"onyi, B.~Szab\'o, G.~T\"or\"os, S.~Wegner, J.~T\"or\"ok, E.~Somfai,
  T.~Bien, and R.~Stannarius.
\newblock Orientational order and alignment of elongated particles induced by
  shear.
\newblock {\em Phys. Rev. Lett.}, 108:228302, 2012.

\bibitem{tamas_review}
T.~B\"orzs\"onyi and R.~Stannarius.
\newblock Granular materials composed of shape-anisotropic grains.
\newblock {\em Soft Matter}, 9:7401, 2013.


\bibitem{Lu2015}
G.~Lu, J.~R. Third, and C.~R. M\"uller.
\newblock Discrete element models for non-spherical particle systems: From
  theoretical developments to applications.
\newblock {\em Chem. Eng. Sci.}, 127:425--465, 2015.


\bibitem{donev_2004}
A.~Donev, I.~Cisse, D.~Sachs, E.A.~Variano, F.H.~Stillinger, R.~Connelly, S.~Torquato, P.M.~Chaikin,
\newblock Improving the density of jammed disordered packings using ellipsoids
\newblock {\em Science}, 303:990, 2004.

\bibitem{frette_1996}
V.~Frette, K.~Christensen, A.~Malthe-S\o renssen, J.~Feder, T.~J\o ssang, and P.~Meakin,
\newblock Avalanche dynamics in a pile of rice.
\newblock {\em Nature}, 379:49, 1996.

\bibitem{narayan_2006}
V.~Narayan, N.~Menon and S.~Ramaswamy,
\newblock  Nonequilibrium steady states in a vibrated-rod monolayer: tetratic, nematic, and smectic correlations.
\newblock  {\em J. Stat. Mech.: Theory Exp.}, 2006:P01005. 2006.

\bibitem{wortel_2015}
G. Wortel, T. B\"orzs\"onyi, E. Somfai, S. Wegner, B. Szab\'o, R. Stannarius, and M. van Hecke,
\newblock Heaping, secondary flows and broken symmetry of elongated granular particles.
\newblock {\em Soft Matter}, 11:2570 (2015) .


\bibitem{Otsu1979}
N.~Otsu.
\newblock Influence of particle shape and surface friction variability on
  response of rod-shaped particulate media.
\newblock {\em IEEE Trans. Systems, Man, Cybernetics}, SMC-9:62, 1979.

\bibitem{wegner_SM_2014}
S.~Wegner, R.~Stannarius, A.~B\"ose, G.~Rose, B.~Szab\'o, E.~Somfai, and
  T.~B\"orzs\"onyi.
\newblock Effects of grain shape on packing and dilatancy of sheared granular
  materials.
\newblock {\em Soft Matter}, 10:5157, 2014.

\bibitem{wegner_2012}
S.~Wegner, T.~B\"orzs\"onyi, T.~Bien, G.~Rose, and R.~Stannarius.
\newblock Alignment and dynamics of elongated cylinders under shear.
\newblock {\em Soft Matter}, 8:10950--10958, 2012.

\bibitem{tamas_2012_pre}
T.~B\"orzs\"onyi, B.~Szab\'o, S.~Wegner, K.~Harth, J.~T\"or\"ok, E.~Somfai,
  T.~Bien, and R.~Stannarius.
\newblock Shear-induced alignment and dynamics of elongated granular particles.
\newblock {\em Phys. Rev. E}, 86:051304, 2012.

\bibitem{szabo_2014_pre}
B.~Szab\'o, J.~T\"or\"ok, E.~Somfai, S.~Wegner, R.~Stannarius, A.~B\"ose,
  G.~Rose, F.~Angenstein, and T.~B\"orzs\"onyi.
\newblock Evolution of shear zones in granular materials.
\newblock {\em Phys. Rev. E}, 90:032205, 2014.

\end{thebibliography}
\end{document}